\documentclass[aps,prl,twocolumn,groupedaddress,superscriptaddress,nobibnotes]{revtex4-2}
\usepackage{graphicx}
\usepackage{epstopdf}
\usepackage{amssymb, amsmath}
\usepackage[usenames]{color}
\usepackage{bm}
\usepackage{ulem}
\usepackage{multirow}
\usepackage{threeparttable}
\usepackage{booktabs}
\usepackage{dcolumn}
\usepackage{float}
\usepackage{hyperref}
\usepackage[columnwise]{lineno}

\hypersetup{
	colorlinks=true,
	linkcolor=blue,
	filecolor=gray,      
	urlcolor=blue,
	citecolor=blue,
}

\date{\today}

\begin{document}
\title{A Catalogue of Topological Moir\'{e} Bands in Twisted Semiconductors}

\author{Jiaheng Li$^\dagger$}
\affiliation{Beijing National Laboratory for Condensed Matter Physics and Institute of Physics, Chinese Academy of Sciences, Beijing 100190, China}

\author{Yan Zhang$^\dagger$}
\affiliation{Beijing National Laboratory for Condensed Matter Physics and Institute of Physics, Chinese Academy of Sciences, Beijing 100190, China}
\affiliation{University of Chinese Academy of Sciences, Beijing 100049, China}

\author{Jiaxuan Liu$^\dagger$}
\affiliation{Beijing National Laboratory for Condensed Matter Physics and Institute of Physics, Chinese Academy of Sciences, Beijing 100190, China}
\affiliation{University of Chinese Academy of Sciences, Beijing 100049, China}

\author{Caiyuan Ye}
\affiliation{Beijing National Laboratory for Condensed Matter Physics and Institute of Physics, Chinese Academy of Sciences, Beijing 100190, China}
\affiliation{University of Chinese Academy of Sciences, Beijing 100049, China}

\author{Tiannian Zhu}
\affiliation{Beijing National Laboratory for Condensed Matter Physics and Institute of Physics, Chinese Academy of Sciences, Beijing 100190, China}
\affiliation{Condensed Matter Physics Data Center of Chinese Academy of Sciences, Beijing 100190,  China}

\author{Zhong Fang}
\affiliation{Beijing National Laboratory for Condensed Matter Physics and Institute of Physics, Chinese Academy of Sciences, Beijing 100190, China}
\affiliation{University of Chinese Academy of Sciences, Beijing 100049, China}

\author{Hongming Weng}
\email{hmweng@iphy.ac.cn}
\affiliation{Beijing National Laboratory for Condensed Matter Physics and Institute of Physics, Chinese Academy of Sciences, Beijing 100190, China}
\affiliation{University of Chinese Academy of Sciences, Beijing 100049, China}
\affiliation{Condensed Matter Physics Data Center of Chinese Academy of Sciences, Beijing 100190,  China}

\author{Quansheng Wu}
\email{quansheng.wu@iphy.ac.cn}
\affiliation{Beijing National Laboratory for Condensed Matter Physics and Institute of Physics, Chinese Academy of Sciences, Beijing 100190, China}
\affiliation{University of Chinese Academy of Sciences, Beijing 100049, China}

\begin{abstract}
Twisted two-dimensional semiconductors provide a route to flat and topological moir\'e minibands, but systematic principles for organizing their material dependence have remained unclear. Here, we establish a high-throughput framework that integrates structural relaxation, first-principles electronic structure calculations, and moir\'e band topology. We apply this framework to 43 experimentally realized monolayers and 91 symmetry-inequivalent bilayer prototypes, yielding over 1,000 angle-resolved moir\'e electronic band structures. This database reveals that the low-energy moir\'e electronic structure is organized primarily by the valley character of the parent band edge together with stacking symmetry. In $\Gamma$-valley systems, the miniband width usually follows a nearly quadratic twist-angle scaling, consistent with a folding-dominated kinetic-energy scale. In $K$-valley systems, stacking-controlled interlayer hybridization governs whether parent Berry curvature is redistributed into isolated valley Chern minibands. By contrast, $M$-valley systems form a more material-specific class associated with anisotropic and symmetry-constrained band folding. The same valley-and-stacking hierarchy rationalizes the emergence or suppression of $\mathbb{Z}_2$ minibands, and surface termination in Janus bilayers provides a microscopic knob for changing the relevant valley character. These results establish a materials-level organizing principle for designing flat and topological moir\'e bands in twisted semiconductors.

\end{abstract}
\maketitle

Twisted layered materials have emerged as a premier platform for engineering quantum matter through rotational misalignment, forming the foundation of twistronics \cite{carr2017twistronics, carr2020electronic, andrei2021marvels, kennes2021moire}. In these systems, moir\'e potentials profoundly renormalize electronic states, enabling a rich landscape of emergent phenomena ranging from correlated insulators and unconventional superconductivity to topological quantum Hall effects \cite{bistritzer2011moire, regnault2011fractional, vizner2021interfacial, jin2019observation, lu2019superconductors, cao2018unconventional, cao2020tunable, wang2020stacking, song2021direct, xie2022twist, yang2023moire, zeng2023thermodynamic, cai2023signatures, xu2023observation, zhang2024polarization, wang2024fractional, redekop2024direct, jia2024moire, xia2025superconductivity, xu2025multiple}. This paradigm, initially centered on graphene \cite{bistritzer2011moire, cao2018unconventional, cao2020tunable}, has rapidly expanded to transition metal dichalcogenides, hexagonal boron nitride, and low-dimensional magnets \cite{jin2019observation, wu2019topological, wang2020stacking, wang2020correlated, xiao2020moire, vizner2021interfacial, song2021direct, mak2022semiconductor, xie2022twist, yang2023moire, zeng2023thermodynamic, cai2023signatures, xu2023observation, zhang2024polarization, wang2024fractional, redekop2024direct, kang2024evidence, jia2024moire, xia2025superconductivity, xu2025multiple, guo2025superconductivity, zhai2025twistronics}. Yet, despite this remarkable versatility, it remains fundamentally unresolved whether universal organizing principles govern moir\'e electronic structures, or whether their behavior is intrinsically material-specific.

The lack of a predictive framework to discern universal scaling from material-dependent evolution stems from the extreme sensitivity of moir\'e bands to parent electronic structures, stacking symmetries, and twist angles \cite{geim2013van, novoselov20162d, carr2017twistronics, andrei2021marvels}. This impasse is largely a consequence of the prohibitive computational cost required to explore the expansive moir\'e parameter space with sufficient fidelity. At small twist angles, supercells containing thousands of atoms render standard first-principles calculations impractical \cite{kresse1996efficient}. Furthermore, the critical role of structural relaxation introduces atomic-scale reconstructions that profoundly reshape band topology, necessitating a level of first-principles precision that is rarely scalable \cite{jia2024moire, zhang2024polarization, xu2025multiple}. Consequently, systematic investigations remain restricted to a few prototypical cases, preventing a global understanding of the interplay between atomic structure and emergent physics.

In parallel, large-scale computational databases have transformed materials discovery by enabling the systematic identification of hidden regularities across high-dimensional spaces \cite{jain2013commentary, curtarolo2013high, kalidindi2015materials, agrawal2016perspective, bradlyn2017topological, haastrup2018computational, mounet2018two, zhang2019catalogue, vergniory2019complete}. However, while recent efforts have initiated the exploration of twistable systems \cite{jia2024moire, zhang2024polarization, zhang2024universal, jiang20242d, xu2025engineer}, these studies typically either survey a broad range of materials at a limited number of twist angles, or focus on specific material families at particular twist configurations, and therefore do not yet offer a systematically comparable dataset across a wide range of twist angles. The absence of such a foundation has hindered the quantitative identification of the organizing principles governing moir\'e electronic structure, particularly the complex interplay between valley character, twist angle, and topology. Addressing this gap requires a comprehensive, high-throughput approach to establish the fundamental rules that dictate the topological and electronic landscape across diverse moir\'e material platforms.

In this work, we bridge this gap by establishing a high-throughput computational framework and constructing a comprehensive database of twisted van der Waals bilayers, which integrates structural relaxation with high-fidelity electronic and topological characterization across an unprecedented range of materials, stacking configurations, and experimentally accessible twist angles. Our systematic analysis uncovers universal organizing principles, most notably a valley-controlled scaling of miniband bandwidths that quantifies the suppression of kinetic energy in moir\'e superlattices. Furthermore, we identify a valley-selective mechanism through which interlayer twisting induces a diverse array of topological minibands from trivial parent systems, stabilizing Chern and $\mathbb{Z}_2$ phases across broad twist-angle windows. These results establish a predictive, quantitative map connecting parent electronic structures to moir\'e-induced topology, providing a rigorous foundation for the rational design and analysis of twisted quantum matter.

\section{Computational workflow}
\begin{figure}[htbp]
	\includegraphics[width=0.95\linewidth]{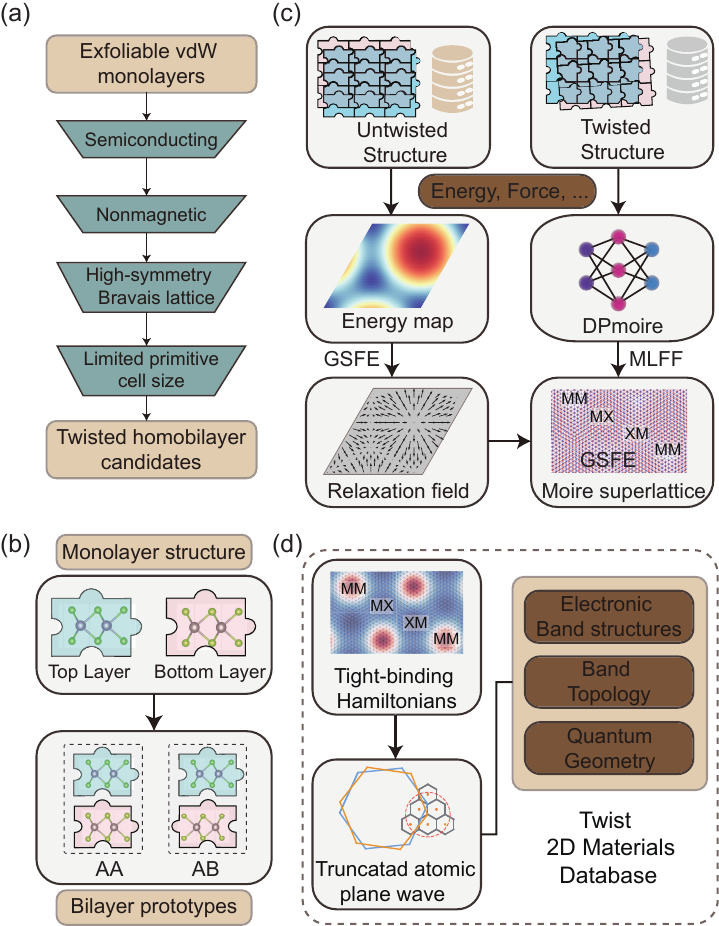}
    \caption{
    \textbf{Hierarchical high-throughput workflow for twisted two-dimensional materials.}
    (\textbf{a}) An experimental 2D materials database serves as the input, from which candidate monolayers are selected using predefined screening criteria.
    (\textbf{b}) The screened monolayers are assembled into symmetry-inequivalent bilayer prototypes by enumerating distinct relative rotations and termination-dependent stacking configurations.
    (\textbf{c}) Interlayer sliding is sampled to construct the generalized stacking-fault energy (GSFE) surface, which provides initial stacking configurations for twisted structures. These are further relaxed using machine-learning force fields (MLFFs).
    (\textbf{d}) Starting from the relaxed structures, corresponding tight-binding Hamiltonians derived from first-principles calculations are constructed. The truncated atomic plane-wave (TAPW) method is then used to compute moir\'e electronic band structures and their topological properties, including band topology and quantum geometry. These results together constitute the twisted two-dimensional materials database.
    }
	\label{fig_workflow}
\end{figure}

To achieve this systematic exploration, we developed an integrated computational workflow organized into several distinct stages, each designed to ensure physical fidelity while  navigating the expansive chemical space of twisted 2D materials (Fig.~\ref{fig_workflow}a).

\paragraph{Monolayer selection and screening.}
The process begins with a curated library of experimentally realized two-dimensional materials \cite{mounet2018two}. To focus on insulating and nonmagnetic systems, we apply a sequence of screening criteria that exclude metallic and magnetic monolayers. We further restrict the scope to materials with square or hexagonal symmetry, which facilitates the exhaustive enumeration of commensurate moir\'e supercells via twist indices \cite{carr2020electronic, naik2022twister}. To maintain computational tractability at small twist angles, we constrain the primitive unit cell size, thereby defining a finalized candidate pool for subsequent bilayer assembly.

\paragraph{Bilayer construction and stacking enumeration.}
The selected monolayers are assembled into bilayer prototypes based on their intrinsic crystal symmetry. Using a group-theoretical approach, we identify all symmetry-inequivalent stacking configurations that arise from relative rotations and surface terminations. For centrosymmetric materials, this analysis identifies parallel (AA) and antiparallel (AB) stackings as the primary prototypes \cite{jia2024moire, zhang2024universal}. In systems lacking mirror-related surfaces, additional configurations are considered to account for distinct termination combinations. For each prototype, we compute the generalized stacking-fault energy (GSFE) surface \cite{carr2018relaxation} by sampling interlayer translations, providing the requisite interlayer potential for the subsequent structural relaxation of twisted moir\'e superlattices.

\begin{figure*}[htbp]
	\includegraphics[width=1.0\linewidth]{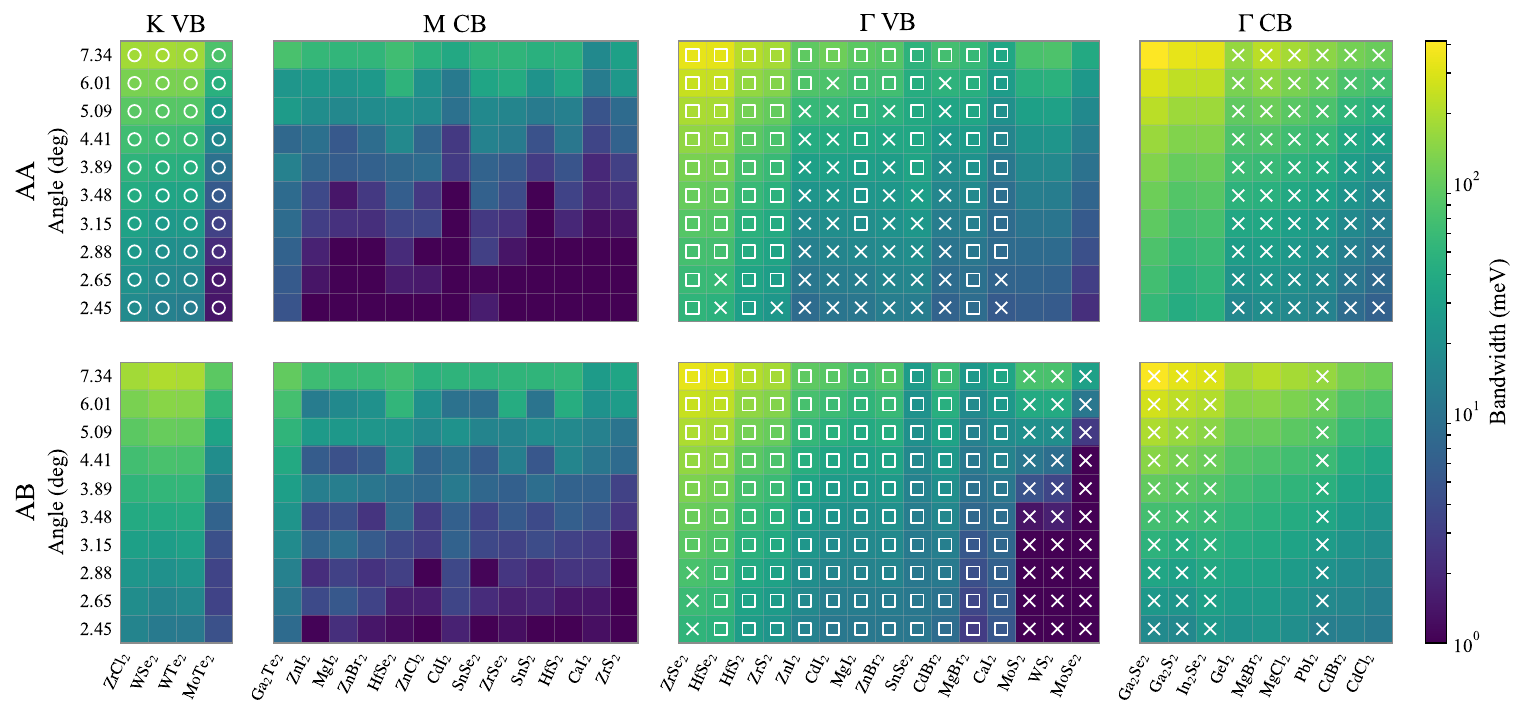}
    \caption{
    \textbf{Twist-angle dependence of moir\'e bandwidth and topological characteristics.} 
    Heatmaps of the bandwidth $W(\theta)$ are presented for various material classes categorized by valley character ($\Gamma$, $K$, and $M$) and band-edge origin. Top and bottom panels display results for the AA- and AB-stacking configurations, respectively. The labels VB and CB represent the moir\'e minibands derived from the parent valence and conduction bands. Color scales encode $W(\theta)$ from the dispersive to the flat regimes. Topological properties are indicated by symbols: squares denote $\mathbb{Z}_2 = 1$ for $\Gamma$-valley systems, circles denote $|C_K| = 1$ for $K$-valley systems, and crosses mark topologically trivial phases. Notably, the absence of symbols identifies minibands that either lack energetic isolation in the $K$- and $\Gamma$-valleys or host sophisticated $M$-valley topology. Governed by emergent non-symmorphic symmetries, these $M$-point systems transcend standard topological classifications and manifest unique kagome or quasi-one-dimensional physics \cite{cualuguaru2025moire, lei2025moire}.
    }
	\label{fig_moire_band_width}
\end{figure*} 

\paragraph{High-accuracy structural relaxation and electronic characterization.}
Twisted structures are relaxed using a combination of GSFE-based continuum relaxation \cite{carr2018relaxation} and E(3)-equivariant machine-learning force fields (MLFFs) \cite{musaelian2023learning}, enabling efficient optimization of large supercells with near first-principles accuracy \cite{liu2025dpmoire}. 
Based on the relaxed geometries, electronic properties are evaluated using density functional theory to construct representative tight-binding Hamiltonians. The low-energy moir\'e minibands and their associated topological invariants, including Chern numbers and $\mathbb{Z}_2$ indices, are subsequently computed using the truncated atomic plane-wave (TAPW) method \cite{miao2023truncated, zhang2024universal, zhang2026tapw_method}. This computational pipeline allows for the high-throughput evaluation of electronic spectra across a wide range of twist angles, culminating in a comprehensive, angle-resolved database of moir\'e band structures, band topology and quantum geometry  (Fig.~\ref{fig_workflow}d). This systematic foundation ultimately enables the identification of the universal organizing principles governing emergent quantum states in semiconductor moir\'e systems.

\section{Results}
Leveraging this framework, we construct an extensive twisted 2D material database by systematically screening 43 experimentally realized monolayers, comprising 39 hexagonal and 4 square lattices. The exhaustive enumeration of symmetry-inequivalent stacking configurations yields 91 bilayer prototypes, whose analysis generates over 1,000 moir\'e band structures. 

\paragraph{Valley classification.}
To distill universal organizing principles from our expansive dataset, we first categorize the triangular-lattice bilayers, which form the largest class in our dataset, based on the valley character of their constituent monolayer band extrema. This classification distinguishes high-symmetry valleys ($\Gamma$, $K$, and $M$) from off-symmetry $Q$ valleys. While systems with $Q$-valley extrema manifest intricate twist-angle dependence, which is driven by the sensitive interplay of moir\'e band folding and inter-valley hybridization, symmetry-anchored valleys exhibit more systematic and class-dependent behaviors. Consequently, our analysis prioritizes these high-symmetry valley systems, as they provide a robust platform for identifying the fundamental rules that govern bandwidth scaling and topological reconstruction in moir\'e materials.

\begin{figure*}[htbp]
	\includegraphics[width=1.0\linewidth]{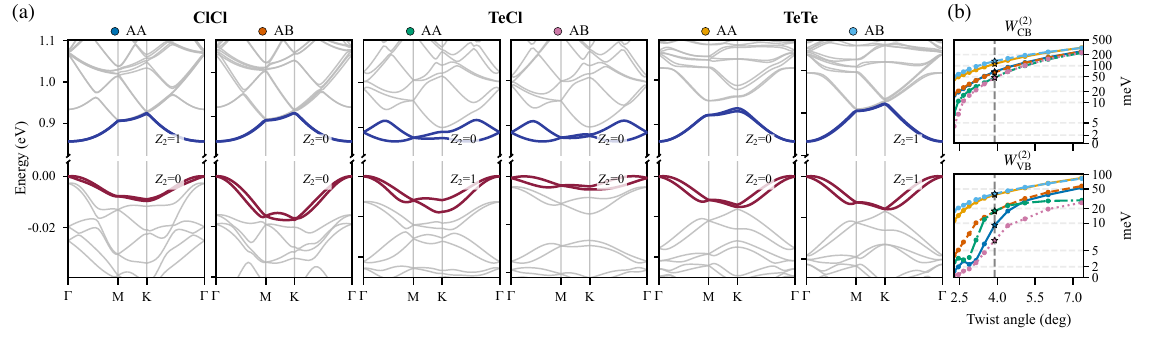}
    \caption{
    \textbf{Stacking-dependent moir\'e band structures in BiTeCl bilayer prototypes.} 
    (a) Electronic band structures at a twist angle of $3.89^\circ$ for representative prototypes, with the $\mathbb{Z}_2$ invariant of the two bands closest to the Fermi level indicated. 
    (b) Twist-angle evolution of the corresponding conduction- and valence-band bandwidths.
    }
	\label{fig_moire_band_BiTeCl}
\end{figure*} 

\paragraph{Universal scaling and valley-controlled hierarchy.}
The global twist-angle dependence of moir\'e bandwidths, visualized as a spectral heatmap in Fig.~\ref{fig_moire_band_width}, reveals a well-defined hierarchy dictated by the valley character of the parent electronic states. A central finding is that the evolution of moir\'e minibands hinges not merely on the high-symmetry momentum of the parent valley, but primarily on the specific valley hosting the low-energy states, which leads to qualitatively distinct regimes of twist-angle dependence. Among these, $\Gamma$-valley systems exhibit the most systematic behavior, where the bandwidth hierarchy across different compounds is largely preserved upon twisting. For most $\Gamma$-valley systems, the bandwidth diminishes smoothly and follows a characteristic quadratic scaling, $W(\theta) \propto \theta^2$, in the small-angle limit, consistent with a folding-dominated regime. Nevertheless, stacking-specific interactions can trigger significant departures from this trend, particularly in AB-stacked transition metal dichalcogenides such as {MoTe$_2$}, where the bandwidth quenches more precipitously \cite{jia2024moire}. Such deviations signal a transition to a strong-coupling regime, in which the large carrier effective mass and the depth of the moir\'e potential cooperatively drive the electronic states toward the atomic limit. Notably, this accelerated quenching is also observed in $K$-valley AB configurations, where spin-prohibited tunneling necessitates a regime of strong potential-induced localization that suppresses the kinetic energy more effectively than in the AA-stacked counterparts \cite{wu2019topological, jia2024moire}. In contrast, $M$-valley systems lack such systematic scaling, as their bandwidth evolution manifests a strong material dependence without a discernible universal trend.

\paragraph{Twist-induced topology from valley structure.}
The valley-dependent hierarchy that governs bandwidth scaling also dictates the nature and robustness of topological minibands (Fig.~\ref{fig_moire_band_width}). In $\Gamma$-valley systems, twisting quenches the kinetic energy, facilitating the formation of well-isolated minibands that are highly susceptible to gap opening and topological reconstruction. The resulting topology is fundamentally dictated by the stacking configuration of the parent untwisted bilayer, a dependence most prominently manifested at the band edges. Specifically, at the valence band maximum (VBM) of systems like CdI$_2$ and HfSe$_2$, one configuration sustains a robust $\mathbb{Z}_2 = 1$ insulating state, whereas another undergoes a transition to a trivial phase ($\mathbb{Z}_2 = 0$) in the small-angle moir\'e regime. Conversely, at the VBM of transition metal dichalcogenides (TMDs) such as MoSe$_2$, AB-stacked moir\'e structures consistently host a trivial topology, while the alternative AA stackings exhibit a vanishingly small gap that mimics the quintessential gapless Kane-Mele model \cite{kane2005topological}. A strikingly parallel stacking dependence is mirrored at the conduction band minimum (CBM), where one configuration is likewise trivial, while the other similarly manifests the gapless Kane-Mele physics. $K$-valley systems offer a complementary realization of symmetry-governed topology, where parent Berry curvature is redistributed among moir\'e minibands. In parallel (AA) stacked bilayers, spin-aligned interlayer tunneling facilitates coherent hybridization, lifting the valley degeneracy and yielding well-isolated minibands with nonzero valley Chern numbers ($|C_K| = 1$). Conversely, antiparallel (AB) stacking preserves inversion symmetry, which, together with spin-valley locking, enforces band degeneracy at $\Gamma$ and typically precludes the isolation of topmost minibands \cite{wu2019topological, jia2024moire, zhang2024universal}. In contrast, $M$-valley systems host a sophisticated topological landscape governed by emergent non-symmorphic symmetries that transcend standard classifications and manifest unique kagome or quasi-one-dimensional physics \cite{lei2025moire, cualuguaru2025moire}. 

Beyond band topology, our database also resolves the quantum geometry of isolated moir\'e manifolds. By quantifying the relative fluctuation $\delta_g$ of the quantum metric $\mathrm{Tr}\,g(\mathbf{k})$, as detailed in Supplementary Section VIII \cite{SM}, we identify geometrically uniform flat topological bands that are more Landau-level-like and therefore more relevant for fractionalized phases \cite{Haldane2011Geometrical,Roy2014BandGeometry,Parameswaran2013FQHFlatBands}. This metric-based screening highlights AA-type MoTe$_2$ among Chern-band systems and several $\mathbb{Z}_2=1$ manifolds, such as AB-stacked BiTeI, MgBr$_2$ and CaI$_2$, as candidates for further many-body investigation.


\paragraph{Surface termination as a microscopic knob for valley engineering.}
The decisive role of valley character is further elucidated through surface termination and stacking configuration, which provide microscopic control over the momentum-space location of band extrema. A quintessential example is found in the Janus homobilayer BiTeCl, where the intrinsic inequivalence of Te and Cl surfaces allows for three distinct bilayer terminations: Cl–Cl, Te–Cl, and Te–Te. Our analysis reveals that these termination environments fundamentally reconfigure the parent valley structure. Specifically, while the symmetric Te–Te and Cl–Cl configurations retain their valence band maxima at the $\Gamma$ point, the asymmetric Te–Cl termination shifts the band extrema to off-symmetry $Q$ valleys \cite{SM}. 

The consequences of this valley shift are directly reflected in the moir\'e spectra at a twist angle of $3.89^\circ$ (Fig.~\ref{fig_moire_band_BiTeCl}). The Te–Cl termination manifests the complex, strongly hybridized dispersions typical of $Q$-valley systems, whereas the symmetric terminations yield well-defined minibands dominated by $\Gamma$-derived folding. Notably, the Te–Te configuration provides a clear demonstration of stacking-controlled topology, where AA stacking results in a trivial phase ($\mathbb{Z}_2=0$), while AB stacking stabilizes a robust $\mathbb{Z}_2=1$ phase over a wide twist-angle range. These findings underscore that surface termination and stacking serve as precise experimental knobs to tune the valley character of parent monolayers, thereby enabling the targeted engineering of moir\'e band reconstruction and emergent topological states.

Complementing these representative examples, we provide a comprehensive dataset in the Supplementary Information \cite{SM}, encompassing moir\'e band structures, gap statistics, and topological characterizations for all investigated materials. To support community-driven discovery, we have developed an interactive online database that enables seamless exploration of this extensive materials library. For each entry in the repository, the platform delivers a detailed profile of the electronic structure and topology, including Wilson loop spectra, Berry curvature maps, and quantum geometry, systematically cross-referenced with the properties of the corresponding parent monolayers. This integrated framework facilitates a direct evaluation of the interplay between moir\'e-induced effects and intrinsic material characteristics, establishing a robust foundation for the design and experimental verification of twisted quantum matter.

\section{Discussion}
In summary, we have established a scalable framework for the systematic exploration of twisted moir\'e semiconductors, uncovering a unified physical landscape that governs their electronic and topological properties across diverse material classes. Our extensive, angle-resolved database reveals that the emergence of moir\'e-induced phenomena is fundamentally dictated by the interplay between valley character and stacking symmetry, which together establish a well-defined hierarchy of bandwidth renormalization and topological reconstruction. Specifically, $\Gamma$-valley systems manifest a near-universal quadratic scaling of bandwidth with twist angle and host robust $\mathbb{Z}_2 = 1$ insulating phases over broad, experimentally accessible regimes. In contrast, $K$-valley systems exhibit valley-resolved topology inherited from the underlying electronic structure, although the manifestation of these phases remains sensitively dependent on the stacking-mediated inversion symmetry. Whereas $M$-valley systems display predominantly idiosyncratic behaviors, characterized by irregular bandwidth evolution and not captured by the standard isolated-band $\mathbb{Z}_2$/Chern classification, the systematic regularity observed in other valleys provides a predictive framework for the rational design of moir\'e matter. Collectively, these findings demonstrate that interlayer twisting, modulated by valley degrees of freedom, serves as a universal mechanism for generating emergent quantum states, laying the foundation for a predictive taxonomy and the targeted engineering of two-dimensional twistronics.

\section{Data availability}
Data supporting the findings of this study are available within the Supplementary Information \cite{SM} and are also publicly accessible at the web-based repository: \url{https://materialsgalaxy.iphy.ac.cn/moire}. The dataset includes relaxed moir\'e structures, electronic band structures, and derived topological quantities such as topological invariants, Wilson-loop spectra, and momentum-resolved Berry curvature and quantum geometry in the Brillouin zone, as well as associated metadata from the twisted materials database.

\section{Acknowledgments}
We are grateful for helpful discussions with J. Yu and B. A. Bernevig. This work was supported by the National Key R\&D Program of China (2024YFA1408400, 2023YFA1607400, 2022YFA1403800), the National Natural Science Foundation of China (Grant Nos. 12274436, 11925408, 11921004), and the Science Center of the National Natural Science Foundation of China (Grant No. 12188101), and the Beijing Municipal Science \& Technology Commission, Administrative Commission of Zhongguancun Science Park No. Z251100003625025. H. Weng acknowledges support from the New Cornerstone Science Foundation (XPLORER PRIZE). J. Li acknowledges support from the China National Postdoctoral Program for Innovative Talents (Grant No. BX20220334).

\section{Contributions}
Q. Wu conceived the study and co-supervised the project with H. Weng. J. Li, Y. Zhang, and J. Liu developed the computational framework and constructed the twisted moir\'e materials database. J. Li screened candidate materials. J. Liu carried out structural relaxations, while Y. Zhang performed first-principles calculations and calculated the electronic and topological properties of large-scale moir\'e structures. Y. Zhang and J. Liu also developed the database website. Y. Cai and T. Zhu assisted with HPC computing and website deployment. J. Li drafted the manuscript with input from all authors. J. Li, Y. Zhang, and J. Liu wrote the Methods section and the Supplementary Information. All authors discussed the results and commented on the manuscript.

\bibliography{reference}

@article{carr2017twistronics,
  title={Twistronics: Manipulating the electronic properties of two-dimensional layered structures through their twist angle},
  author={Carr, Stephen and Massatt, Daniel and Fang, Shiang and Cazeaux, Paul and Luskin, Mitchell and Kaxiras, Efthimios},
  journal={Phys. Rev. B},
  volume={95},
  number={7},
  pages={075420},
  year={2017},
  url={https://journals.aps.org/prb/abstract/10.1103/PhysRevB.95.075420},
  publisher={APS}
}

@article{carr2020electronic,
  title={Electronic-structure methods for twisted moir{\'e} layers},
  author={Carr, Stephen and Fang, Shiang and Kaxiras, Efthimios},
  journal={Nat. Rev. Mater.},
  volume={5},
  number={10},
  pages={748--763},
  year={2020},
  url={https://www.nature.com/articles/s41578-020-0214-0},
  publisher={Nature Publishing Group UK London}
}

@article{andrei2021marvels,
  title={The marvels of moir{\'e} materials},
  author={Andrei, Eva Y and Efetov, Dmitri K and Jarillo-Herrero, Pablo and MacDonald, Allan H and Mak, Kin Fai and Senthil, T and Tutuc, Emanuel and Yazdani, Ali and Young, Andrea F},
  journal={Nat. Rev. Mater.},
  volume={6},
  number={3},
  pages={201--206},
  year={2021},
  url={https://www.nature.com/articles/s41578-021-00284-1},
  publisher={Nature Publishing Group UK London}
}

@article{kennes2021moire,
  title={Moir{\'e} heterostructures as a condensed-matter quantum simulator},
  author={Kennes, Dante M and Claassen, Martin and Xian, Lede and Georges, Antoine and Millis, Andrew J and Hone, James and Dean, Cory R and Basov, DN and Pasupathy, Abhay N and Rubio, Angel},
  journal={Nat. Phys.},
  volume={17},
  number={2},
  pages={155--163},
  year={2021},
  url={https://www.nature.com/articles/s41567-020-01154-3},
  publisher={Nature Publishing Group UK London}
}

@article{bistritzer2011moire,
author = {Rafi Bistritzer  and Allan H. MacDonald },
title = {Moiré bands in twisted double-layer graphene},
journal = {Proc. Natl. Acad. Sci.},
volume = {108},
number = {30},
pages = {12233-12237},
year = {2011},
doi = {10.1073/pnas.1108174108},
URL = {https://www.pnas.org/doi/abs/10.1073/pnas.1108174108},
}

@article{regnault2011fractional,
  title = {Fractional {Chern} Insulator},
  author = {Regnault, N. and Bernevig, B. Andrei},
  journal = {Phys. Rev. X},
  volume = {1},
  issue = {2},
  pages = {021014},
  numpages = {14},
  year = {2011},
  month = {Dec},
  publisher = {American Physical Society},
  doi = {10.1103/PhysRevX.1.021014},
  url = {https://link.aps.org/doi/10.1103/PhysRevX.1.021014}
}

@article{vizner2021interfacial,
  title={Interfacial ferroelectricity by van der {Waals} sliding},
  author={Vizner Stern, Maayan and Waschitz, Yuval and Cao, Wei and Nevo, Iftach and Watanabe, Kenji and Taniguchi, Takashi and Sela, Eran and Urbakh, Michael and Hod, Oded and Ben Shalom, Moshe},
  journal={Science},
  volume={372},
  number={6549},
  pages={1462--1466},
  year={2021},
  url={https://www.science.org/doi/10.1126/science.abe8177},
  publisher={American Association for the Advancement of Science}
}

@article{jin2019observation,
  title={Observation of moir{\'e} excitons in {WSe$_2$/WS$_2$} heterostructure superlattices},
  author={Jin, Chenhao and Regan, Emma C and Yan, Aiming and Iqbal Bakti Utama, M and Wang, Danqing and Zhao, Sihan and Qin, Ying and Yang, Sijie and Zheng, Zhiren and Shi, Shenyang and others},
  journal={Nature},
  volume={567},
  number={7746},
  pages={76--80},
  year={2019},
  url={https://www.nature.com/articles/s41586-019-0976-y},
  publisher={Nature Publishing Group UK London}
}

@article{lu2019superconductors,
  title={Superconductors, orbital magnets and correlated states in magic-angle bilayer graphene},
  author={Lu, Xiaobo and Stepanov, Petr and Yang, Wei and Xie, Ming and Aamir, Mohammed Ali and Das, Ipsita and Urgell, Carles and Watanabe, Kenji and Taniguchi, Takashi and Zhang, Guangyu and others},
  journal={Nature},
  volume={574},
  number={7780},
  pages={653--657},
  year={2019},
  publisher={Nature Publishing Group UK London}
}

@article{cao2020tunable,
  title={Tunable correlated states and spin-polarized phases in twisted bilayer--bilayer graphene},
  author={Cao, Yuan and Rodan-Legrain, Daniel and Rubies-Bigorda, Oriol and Park, Jeong Min and Watanabe, Kenji and Taniguchi, Takashi and Jarillo-Herrero, Pablo},
  journal={Nature},
  volume={583},
  number={7815},
  pages={215--220},
  year={2020},
  url={https://www.nature.com/articles/s41586-020-2260-6},
  publisher={Nature Publishing Group UK London}
}

@article{cao2018unconventional,
  title={Unconventional superconductivity in magic-angle graphene superlattices},
  author={Cao, Yuan and Fatemi, Valla and Fang, Shiang partners and Watanabe, Kenji and Taniguchi, Takashi and Kaxiras, Efthimios and Jarillo-Herrero, Pablo},
  journal={Nature},
  volume={556},
  number={7699},
  pages={43--50},
  year={2018},
  publisher={Nature Publishing Group UK London},
  url={https://www.nature.com/articles/nature26160}
}

@article{wang2020stacking,
  title={Stacking domain wall magnons in twisted van der {Waals} magnets},
  author={Wang, Chong and Gao, Yuan and Lv, Hongyan and Xu, Xiaodong and Xiao, Di},
  journal={Phys. Rev. Lett.},
  volume={125},
  number={24},
  pages={247201},
  year={2020},
  publisher={APS}
}

@article{song2021direct,
  title={Direct visualization of magnetic domains and moir{\'e} magnetism in twisted {2D} magnets},
  author={Song, Tiancheng and Sun, Qi-Chao and Anderson, Eric and Wang, Chong and Qian, Jimin and Taniguchi, Takashi and Watanabe, Kenji and McGuire, Michael A and St{\"o}hr, Rainer and Xiao, Di and others},
  journal={Science},
  volume={374},
  number={6571},
  pages={1140--1144},
  year={2021},
  url={https://www.science.org/doi/full/10.1126/science.abj7478},
  publisher={American Association for the Advancement of Science}
}

@article{xie2022twist,
  title={Twist engineering of the two-dimensional magnetism in double bilayer chromium triiodide homostructures},
  author={Xie, Hongchao and Luo, Xiangpeng and Ye, Gaihua and Ye, Zhipeng and Ge, Haiwen and Sung, Suk Hyun and Rennich, Emily and Yan, Shaohua and Fu, Yang and Tian, Shangjie and others},
  journal={Nat. Phys.},
  volume={18},
  number={1},
  pages={30--36},
  year={2022},
  url={https://www.nature.com/articles/s41567-021-01408-8},
  publisher={Nature Publishing Group UK London}
}

@article{yang2023moire,
  title={Moir{\'e} magnetic exchange interactions in twisted magnets},
  author={Yang, Baishun and Li, Yang and Xiang, Hongjun and Lin, Haiqing and Huang, Bing},
  journal={Nat. Comput. Sci.},
  volume={3},
  number={4},
  pages={314--320},
  year={2023},
  url={https://www.nature.com/articles/s43588-023-00430-5},
  publisher={Nature Publishing Group US New York}
}

@article{zeng2023thermodynamic,
  title={Thermodynamic evidence of fractional {Chern} insulator in moir{\'e} {MoTe$_2$}},
  author={Zeng, Yihang and Xia, Zhengchao and Kang, Kaifei and Zhu, Jiacheng and Kn{\"u}ppel, Patrick and Vaswani, Chirag and Watanabe, Kenji and Taniguchi, Takashi and Mak, Kin Fai and Shan, Jie},
  journal={Nature},
  volume={622},
  number={7981},
  pages={69--73},
  year={2023},
  url={https://www.nature.com/articles/s41586-023-06452-3},
  publisher={Nature Publishing Group UK London}
}

@article{zhang2024polarization,
  title={Polarization-driven band topology evolution in twisted {MoTe$_2$} and {WSe$_2$}},
  author={Zhang, Xiao-Wei and Wang, Chong and Liu, Xiaoyu and Fan, Yueyao and Cao, Ting and Xiao, Di},
  journal={Nat. Commun.},
  volume={15},
  number={1},
  pages={4223},
  year={2024},
  doi={https://www.nature.com/articles/s41467-024-48511-x},
  publisher={Nature Publishing Group UK London}
}

@article{wang2024fractional,
  title={Fractional {Chern} insulator in twisted bilayer {MoTe$_2$}},
  author={Wang, Chong and Zhang, Xiao-Wei and Liu, Xiaoyu and He, Yuchi and Xu, Xiaodong and Ran, Ying and Cao, Ting and Xiao, Di},
  journal={Phys. Rev. Lett.},
  volume={132},
  number={3},
  pages={036501},
  year={2024},
  url={https://journals.aps.org/prl/abstract/10.1103/PhysRevLett.132.036501},
  publisher={APS}
}

@article{redekop2024direct,
  title={Direct magnetic imaging of fractional {Chern} insulators in twisted {MoTe$_2$}},
  author={Redekop, Evgeny and Zhang, Canxun and Park, Heonjoon and Cai, Jiaqi and Anderson, Eric and Sheekey, Owen and Arp, Trevor and Babikyan, Grigory and Salters, Samuel and Watanabe, Kenji and others},
  journal={Nature},
  volume={635},
  number={8039},
  pages={584--589},
  year={2024},
  url={https://www.nature.com/articles/s41586-024-08153-x},
  publisher={Nature Publishing Group UK London}
}

@article{jia2024moire,
  title={Moir{\'e} fractional {Chern} insulators. {I}. {First}-principles calculations and continuum models of twisted bilayer {MoTe$_2$}},
  author={Jia, Yujin and Yu, Jiabin and Liu, Jiaxuan and Herzog-Arbeitman, Jonah and Qi, Ziyue and Pi, Hanqi and Regnault, Nicolas and Weng, Hongming and Bernevig, B Andrei and Wu, Quansheng},
  journal={Phys. Rev. B},
  volume={109},
  number={20},
  pages={205121},
  year={2024},
  url={https://journals.aps.org/prb/abstract/10.1103/PhysRevB.109.205121},
  publisher={APS}
}

@article{xia2025superconductivity,
  title={Superconductivity in twisted bilayer {WSe$_2$}},
  author={Xia, Yiyu and Han, Zhongdong and Watanabe, Kenji and Taniguchi, Takashi and Shan, Jie and Mak, Kin Fai},
  journal={Nature},
  volume={637},
  number={8047},
  pages={833--838},
  year={2025},
  url={https://www.nature.com/articles/s41586-024-08116-2},
  publisher={Nature Publishing Group UK London}
}

@article{xu2025multiple,
  title={Multiple {Chern} bands in twisted {MoTe$_2$} and possible {non-Abelian} states},
  author={Xu, Cheng and Mao, Ning and Zeng, Tiansheng and Zhang, Yang},
  journal={Phys. Rev. Lett.},
  volume={134},
  number={6},
  pages={066601},
  year={2025},
  url={https://journals.aps.org/prl/abstract/10.1103/PhysRevLett.134.066601},
  publisher={APS}
}

@article{wu2019topological,
  title={Topological insulators in twisted transition metal dichalcogenide homobilayers},
  author={Wu, Fengcheng and Lovorn, Timothy and Tutuc, Emanuel and Martin, Ivar and MacDonald, AH},
  journal={Phys. Rev. Lett.},
  volume={122},
  number={8},
  pages={086402},
  year={2019},
  publisher={APS}
}

@article{wang2020correlated,
  title={Correlated electronic phases in twisted bilayer transition metal dichalcogenides},
  author={Wang, Lei and Shih, En-Min and Ghiotto, Augusto and Xian, Lede and Rhodes, Daniel A and Tan, Cheng and Claassen, Martin and Kennes, Dante M and Bai, Yusong and Kim, Bumho and others},
  journal={Nat. Mater.},
  volume={19},
  number={8},
  pages={861--866},
  year={2020},
  publisher={Nature Publishing Group UK London}
}

@article{xiao2020moire,
  title={Moir{\'e} is more: access to new properties of two-dimensional layered materials},
  author={Xiao, Yao and Liu, Jinglu and Fu, Lei},
  journal={Matter},
  volume={3},
  number={4},
  pages={1142--1161},
  year={2020},
  publisher={Elsevier}
}

@article{mak2022semiconductor,
  title={Semiconductor moir{\'e} materials},
  author={Mak, Kin Fai and Shan, Jie},
  journal={Nat. Nanotech.},
  volume={17},
  number={7},
  pages={686--695},
  year={2022},
  publisher={Nature Publishing Group UK London}
}

@article{kang2024evidence,
  title={Evidence of the fractional quantum spin {Hall} effect in moir{\'e} {MoTe$_2$}},
  author={Kang, Kaifei and Shen, Bowen and Qiu, Yichen and Zeng, Yihang and Xia, Zhengchao and Watanabe, Kenji and Taniguchi, Takashi and Shan, Jie and Mak, Kin Fai},
  journal={Nature},
  volume={628},
  number={8008},
  pages={522--526},
  year={2024},
  publisher={Nature Publishing Group UK London}
}

@article{guo2025superconductivity,
  title={Superconductivity in 5.0° twisted bilayer {WSe$_2$}},
  author={Guo, Yinjie and Pack, Jordan and Swann, Joshua and Holtzman, Luke and Cothrine, Matthew and Watanabe, Kenji and Taniguchi, Takashi and Mandrus, David G and Barmak, Katayun and Hone, James and others},
  journal={Nature},
  volume={637},
  number={8047},
  pages={839--845},
  year={2025},
  url={https://www.nature.com/articles/s41586-024-08381-1},
  publisher={Nature Publishing Group UK London}
}

@article{zhai2025twistronics,
  title={Twistronics and moir{\'e} superlattice physics in {2D} transition metal dichalcogenides},
  author={Zhai, Dawei and Yu, Hongyi and Yao, Wang},
  journal={Rep. Prog. Phys.},
  volume={88},
  number={8},
  pages={084501},
  year={2025},
  publisher={IOP Publishing}
}

@article{geim2013van,
  title={Van der {Waals} heterostructures},
  author={Geim, Andre K and Grigorieva, Irina V},
  journal={Nature},
  volume={499},
  number={7459},
  pages={419--425},
  year={2013},
  publisher={Nature Publishing Group UK London}
}

@article{jain2013commentary,
  title={Commentary: {The Materials Project}: A materials genome approach to accelerating materials innovation},
  author={Jain, Anubhav and Ong, Shyue Ping and Hautier, Geoffroy and Chen, Wei and Richards, William Davidson and Dacek, Stephen and Cholia, Shreyas and Gunter, Dan and Skinner, David and Ceder, Gerbrand and others},
  journal={APL Mater.},
  volume={1},
  number={1},
  pages={011002},
  year={2013},
  url={https://pubs.aip.org/aip/apm/article/1/1/011002/119685},
  publisher={AIP Publishing}
}

@article{kresse1996efficient,
  title = {Efficient iterative schemes for ab initio total-energy calculations using a plane-wave basis set},
  author = {Kresse, G. and Furthm\"uller, J.},
  journal = {Phys. Rev. B},
  volume = {54},
  issue = {16},
  pages = {11169--11186},
  numpages = {0},
  year = {1996},
  month = {Oct},
  publisher = {American Physical Society},
  doi = {10.1103/PhysRevB.54.11169},
  url = {https://link.aps.org/doi/10.1103/PhysRevB.54.11169}
}

@article{novoselov20162d,
  title={2D materials and van der {Waals} heterostructures},
  author={Novoselov, K S and Mishchenko, Artem and Carvalho, Alexandra and Castro Neto, AH},
  journal={Science},
  volume={353},
  number={6298},
  pages={aac9439},
  year={2016},
  publisher={American Association for the Advancement of Science}
}

@article{curtarolo2013high,
  title={The high-throughput highway to computational materials design},
  author={Curtarolo, Stefano and Hart, Gus LW and Nardelli, Marco Buongiorno and Mingo, Natalio and Sanvito, Stefano and Levy, Ohad},
  journal={Nat. Mater.},
  volume={12},
  number={3},
  pages={191--201},
  year={2013},
  publisher={Nature Publishing Group UK London}
}

@article{kalidindi2015materials,
  title={Materials data science: current status and future outlook},
  author={Kalidindi, Surya R and De Graef, Marc},
  journal={Annu. Rev. Mater. Res.},
  volume={45},
  number={1},
  pages={171--193},
  year={2015},
  publisher={Annual Reviews}
}

@article{agrawal2016perspective,
  title={Perspective: Materials informatics and big data: Realization of the “fourth paradigm” of science in materials science},
  author={Agrawal, Ankit and Choudhary, Alok},
  journal={APL Mater.},
  volume={4},
  number={5},
  year={2016},
  publisher={AIP Publishing}
}

@article{bradlyn2017topological,
  title={Topological quantum chemistry},
  author={Bradlyn, Barry and Elcoro, Luis and Cano, Jennifer and Vergniory, Maia G and Wang, Zhijun and Felser, Claudia and Aroyo, Mois I and Bernevig, B Andrei},
  journal={Nature},
  volume={547},
  number={7663},
  pages={298--305},
  year={2017},
  url={https://www.nature.com/articles/nature23268},
  publisher={Nature Publishing Group UK London}
}

@article{haastrup2018computational,
  title={The {Computational 2D Materials Database}: high-throughput modeling and discovery of atomically thin crystals},
  author={Haastrup, Sten and Strange, Mikkel and Pandey, Mohnish and Deilmann, Thorsten and Schmidt, Per S and Hinsche, Nicki F and Gjerding, Morten N and Torelli, Daniele and Larsen, Peter M and Riis-Jensen, Anders C and others},
  journal={2D Mater.},
  volume={5},
  number={4},
  pages={042002},
  year={2018},
  url={https://iopscience.iop.org/article/10.1088/2053-1583/aacfc1},
  publisher={IOP Publishing}
}

@article{mounet2018two,
  title={Two-dimensional materials from high-throughput computational exfoliation of experimentally known compounds},
  author={Mounet, Nicolas and Gibertini, Marco and Schwaller, Philippe and Campi, Davide and Merkys, Andrius and Marrazzo, Antimo and Sohier, Thibault and Castelli, Ivano Eligio and Cepellotti, Andrea and Pizzi, Giovanni and others},
  journal={Nat. Nanotech.},
  volume={13},
  number={3},
  pages={246--252},
  year={2018},
  url={https://www.nature.com/articles/s41565-017-0035-5},
  publisher={Nature Publishing Group UK London}
}

@article{zhang2019catalogue,
  title={Catalogue of topological electronic materials},
  author={Zhang, Tiantian and Jiang, Yi and Song, Zhida and Huang, He and He, Yuqing and Fang, Zhong and Weng, Hongming and Fang, Chen},
  journal={Nature},
  volume={566},
  number={7745},
  pages={475--479},
  year={2019},
  url={https://www.nature.com/articles/s41586-019-0944-6},
  publisher={Nature Publishing Group UK London}
}

@article{vergniory2019complete,
  title={A complete catalogue of high-quality topological materials},
  author={Vergniory, MG and Elcoro, L and Felser, Claudia and Regnault, Nicolas and Bernevig, B Andrei and Wang, Zhijun},
  journal={Nature},
  volume={566},
  number={7745},
  pages={480--485},
  year={2019},
  url={https://www.nature.com/articles/s41586-019-0954-4},
  publisher={Nature Publishing Group UK London}
}

@article{jiang20242d,
  title={2D theoretically twistable material database},
  author={Jiang, Yi and Petralanda, Urko and Skorupskii, Grigorii and Xu, Qiaoling and Pi, Hanqi and C{\u{a}}lug{\u{a}}ru, Dumitru and Hu, Haoyu and Xie, Jiaze and Mustaf, Rose Albu and H{\"o}hn, Peter and others},
  journal={arXiv preprint arXiv:2411.09741},
  year={2024}
}

@article{xu2025engineer,
  title = {Engineering 2D Square Lattice {Hubbard} Models in 90\ifmmode^\circ\else\textdegree\fi{} Twisted $\mathrm{GeX}/\mathrm{SnX}$ ($\mathrm{X}=\mathrm{S}$, Se) Moir\'e Superlattices},
  author = {Xu, Qiaoling and Fischer, Ammon and Tancogne-Dejean, Nicolas and Zhang, Tao and Bostr\"om, Emil Vi\~nas and Claassen, Martin and Kennes, Dante M. and Rubio, Angel and Xian, Lede},
  journal = {Phys. Rev. X},
  volume = {15},
  issue = {4},
  pages = {041049},
  numpages = {18},
  year = {2025},
  month = {Dec},
  publisher = {American Physical Society},
  doi = {10.1103/wcbz-lbr1},
  url = {https://link.aps.org/doi/10.1103/wcbz-lbr1}
}

@article{naik2022twister,
  title={Twister: Construction and structural relaxation of commensurate moir{\'e} superlattices},
  author={Naik, Saismit and Naik, Mit H and Maity, Indrajit and Jain, Manish},
  journal={Comput. Phys. Commun.},
  volume={271},
  pages={108184},
  year={2022},
  publisher={Elsevier}
}

@article{musaelian2023learning,
  title = {Learning local equivariant representations for large-scale atomistic dynamics},
  author = {Musaelian, Albert and Batzner, Simon and Johansson, Anders and Sun, Lixin and Owen, Cameron J and Kornbluth, Mordechai and Kozinsky, Boris},
  journal = {Nat. Commun.},
  volume = {14},
  number = {1},
  pages = {579},
  year = {2023}
}

@article{cualuguaru2025moire,
  title={Moir{\'e} materials based on {M}-point twisting},
  author={C{\u{a}}lug{\u{a}}ru, Dumitru and Jiang, Yi and Hu, Haoyu and Pi, Hanqi and Yu, Jiabin and Vergniory, Maia G and Shan, Jie and Felser, Claudia and Schoop, Leslie M and Efetov, Dmitri K and others},
  journal={Nature},
  volume={643},
  number={8071},
  pages={376--381},
  year={2025},
  publisher={Nature Publishing Group UK London}
}

@article{lei2025moire,
  title={Moir{\'e} band theory for {M}-valley twisted transition metal dichalcogenides},
  author={Lei, Chao and Mahon, Perry T and MacDonald, Allan H},
  journal={Phys. Rev. Lett.},
  volume={135},
  number={19},
  pages={196402},
  year={2025},
  publisher={APS}
}

@article{kane2005topological,
  title = {${Z}_{2}$ Topological Order and the Quantum Spin {Hall} Effect},
  author = {Kane, C. L. and Mele, E. J.},
  journal = {Phys. Rev. Lett.},
  volume = {95},
  issue = {14},
  pages = {146802},
  numpages = {4},
  year = {2005},
  month = {Sep},
  publisher = {American Physical Society},
  doi = {10.1103/PhysRevLett.95.146802},
  url = {https://link.aps.org/doi/10.1103/PhysRevLett.95.146802}
}

@misc{SM,
	note={See Supplementary Information for details on the computational framework, as well as the electronic structures and energy landscapes of screened parent materials in monolayer, bilayer, and bulk forms. It also provides a comprehensive analysis of the electronic and topological properties of moir\'{e} superlattices across a wide range of experimentally feasible twist angles.}
}

@article{carr2018relaxation,
  title = {Relaxation and domain formation in incommensurate two-dimensional heterostructures},
  author = {Carr, Stephen and Massatt, Daniel and Torrisi, Steven B and Cazeaux, Paul and Luskin, Mitchell and Kaxiras, Efthimios},
  journal = {Phys. Rev. B},
  volume = {98},
  number = {22},
  pages = {224102},
  year = {2018},
  url = {https://journals.aps.org/prb/abstract/10.1103/PhysRevB.98.224102}
}

@article{liu2025dpmoire,
  title = {{DPmoire}: a tool for constructing accurate machine learning force fields in moir{\'e} systems},
  author = {Liu, Jiaxuan and Fang, Zhong and Weng, Hongming and Wu, Quansheng},
  journal = {npj Comput. Mater.},
  volume = {11},
  number = {1},
  pages = {248},
  year = {2025},
  url = {https://www.nature.com/articles/s41524-025-01740-0}
}

@article{miao2023truncated,
  title = {Truncated atomic plane wave method for subband structure calculations of moir{\'e} systems},
  author = {Miao, Wangqian and Li, Chu and Han, Xu and Pan, Ding and Dai, Xi},
  journal = {Phys. Rev. B},
  volume = {107},
  number = {12},
  pages = {125112},
  year = {2023}
}

@article{zhang2024universal,
  title = {Universal moir{\'e}-Model-Building Method without Fitting: Application to Twisted {MoTe$_2$} and {WSe$_2$}},
  author = {Zhang, Yan and Pi, Hanqi and Liu, Jiaxuan and Miao, Wangqian and Qi, Ziyue and Regnault, Nicolas and Weng, Hongming and Dai, Xi and Bernevig, B Andrei and Wu, Quansheng and others},
  journal = {arXiv:2411.08108},
  year = {2024}
}

@article{Haldane2011Geometrical,
  author = {Haldane, F. D. M.},
  title = {Geometrical Description of the Fractional Quantum {Hall} Effect},
  journal = {Phys. Rev. Lett.},
  volume = {107},
  pages = {116801},
  year = {2011},
  doi = {10.1103/PhysRevLett.107.116801}
}

@article{Roy2014BandGeometry,
  author = {Roy, Rahul},
  title = {Band geometry of fractional topological insulators},
  journal = {Phys. Rev. B},
  volume = {90},
  pages = {165139},
  year = {2014},
  doi = {10.1103/PhysRevB.90.165139}
}

@article{Parameswaran2013FQHFlatBands,
  author = {Parameswaran, Siddharth A. and Roy, Rahul and Sondhi, S. L.},
  title = {Fractional quantum {Hall} physics in topological flat bands},
  journal = {C. R. Physique},
  volume = {14},
  number = {9--10},
  pages = {816--839},
  year = {2013},
  doi = {10.1016/j.crhy.2013.04.003}
}

@article{xu2023observation,
  title={Observation of integer and fractional quantum anomalous {Hall} effects in twisted bilayer {MoTe$_2$}},
  author={Xu, Fan and Sun, Zheng and Jia, Tongtong and Liu, Chang and Xu, Cheng and Li, Chushan and Gu, Yu and Watanabe, Kenji and Taniguchi, Takashi and Tong, Bingbing and others},
  journal={Phys. Rev. X},
  volume={13},
  number={3},
  pages={031037},
  year={2023},
  url={https://journals.aps.org/prx/abstract/10.1103/PhysRevX.13.031037},
  publisher={APS}
}

@article{cai2023signatures,
  title={Signatures of fractional quantum anomalous {Hall} states in twisted {MoTe$_2$}},
  author={Cai, Jiaqi and Anderson, Eric and Wang, Chong and Zhang, Xiaowei and Liu, Xiaoyu and Holtzmann, William and Zhang, Yinong and Fan, Fengren and Taniguchi, Takashi and Watanabe, Kenji and others},
  journal={Nature},
  volume={622},
  number={7981},
  pages={63--68},
  year={2023},
  url={https://www.nature.com/articles/s41586-023-06289-w},
  publisher={Nature Publishing Group UK London}
}

@misc{zhang2026tapw_method,
	note={Y. Zhang, Q. Wu et al., In preparation (2026).}
}
\end{document}